\documentclass[conference]{IEEEtran}
\IEEEoverridecommandlockouts
\usepackage{cite}
\usepackage{amsmath,amssymb,amsfonts}
\usepackage{algorithmic}
\usepackage{graphicx}
\usepackage{ragged2e}
\usepackage{blindtext}
\usepackage{textcomp}
\usepackage{tabularx}
\usepackage[font=small,labelfont=bf]{caption}
\usepackage{xcolor}
\usepackage{floatrow}
\DeclareFloatFont{footnotezise}{\footnotesize}
\def\BibTeX{{\rm B\kern-.05em{\sc i\kern-.025em b}\kern-.08em
    T\kern-.1667em\lower.7ex\hbox{E}\kern-.125emX}}

\newcommand{\beqn}{\begin{eqnarray}}
\newcommand{\eeqn}{\end{eqnarray}}

\begin{document}

\title{Analyze the Effects of COVID-19 on Energy Storage Systems: A Techno-Economic Approach
\thanks{Nhat Le, Alexis Plasencia Leos, and Juan Henriquez are undergraduate students at the Electrical and Computer Engineering Department of North Carolina Agricultural and Technical State University. This research is supported by Intel Corporation under the HBCU (Historically Black Colleges and Universities)  undergraduate fellowship program.}
}
\author{\IEEEauthorblockN{{Nhat Le, Alexis Plasencia Leos, Juan Henriquez, Anh Phuong Ngo, Hieu T. Nguyen}}
\IEEEauthorblockN{\textit {Department of Electrical \& Computer Engineering,
North Carolina A\&T State University}\\
\{\textit{nmle,~aplasencialeos,~jchenriquez,~ango1\}@aggies.ncat.edu, htnguyen1@ncat.edu}}
}
\maketitle

\begin{abstract}
During the COVID-19 pandemic, the U.S. power sector witnessed remarkable electricity demand changes in many geographical regions. 
These changes were evident in population-dense cities.
This paper incorporates a techno-economic analysis of energy storage systems (ESSs) to investigate the pandemic’s influence on ESS development.
In particular,  we employ a linear program-based revenue maximization model to capture the revenues of ESS from participating in the electricity market, by performing arbitrage on the energy trading, and regulation market, by providing regulation services to stabilize the grid's frequency. 
We consider five dominant energy storage technologies in the U.S., namely, Lithium-ion, Advanced Lead Acid, Flywheel, Vanadium Redox Flow, and Lithium-Iron Phosphate storage technologies.
Extensive numerical results conducted on the case of New York City (NYC) allow us to highlight the negative impact that COVID-19 had on the NYC power sector.
\end{abstract}

\begin{IEEEkeywords}
Energy storage, arbitrage and regulation services, linear programming, COVID-19 effect.
\end{IEEEkeywords}

{ 
\footnotesize
\section*{Nomenclature}
\addcontentsline{toc}{section}{Nomenclature}
\subsection{Set and Indices}
\begin{IEEEdescription}[\IEEEusemathlabelsep\IEEEsetlabelwidth{$V_1,V_2,V_3$}]
\raggedright 
\item[\rule{0pt}{5pt}$\mathcal{T}$,$t$] Time set and time steps, $t\in\mathcal{T}$
\end{IEEEdescription}
\subsection{Parameters}
\begin{IEEEdescription}[\IEEEusemathlabelsep\IEEEsetlabelwidth{$V_1,V_2,V_3$}]
\raggedright 
\item[\rule{0pt}{5pt}$\overline{Q}$] Power rating [MWh]
\item[\rule{0pt}{5pt}$\overline{S}$] Energy capacity [MWh]
\item[\rule{0pt}{5pt}$\eta_{s}$] Self-discharge efficiency [\%]
\item[\rule{0pt}{5pt}$\eta_{c}$] Round-trip efficiency [\%]
\item[\rule{0pt}{5pt}$\mathcal{R}$] Discount/interest rate [\%]
\item[\rule{0pt}{5pt}$\delta_{t}^{ru}$] Fraction of regulation up provided from time period $t$ to $t$+1 [\%]
\item[\rule{0pt}{5pt}$\delta_{t}^{rd}$] Fraction of regulation down provided from time period $t$ to $t$+1 [\%]
\item[\rule{0pt}{5pt}$\gamma_{t}$] Performance score from time period $t$ to $t$+1 [--]
\item[\rule{0pt}{5pt}$\beta_{t}$] Mileage ratio from time period $t$ to $t$+1 [--]
\item[\rule{0pt}{5pt}$\lambda_{t}$] Electricity price from time period $t$ to $t$+1 [\$/MWh]
\item[\rule{0pt}{5pt}$\lambda_{t}^{c}$] Freq. regulation capacity price from time period $t$ to $t$+1 [\$/MWh]
\item[\rule{0pt}{5pt}$\lambda_{t}^{p}$] Freq. regulation movement (performance) price from time $t$ to $t$+1 [\$/MWh]
\end{IEEEdescription}
\subsection{Variables}
\begin{IEEEdescription}[\IEEEusemathlabelsep\IEEEsetlabelwidth{$V_1,V_2,V_3$}]
\raggedright
\item[$q_{t}^{r}$]  Energy charged from time period $t$ to $t$+1 [MWh]
\item[\rule{0pt}{5pt}$q_{t}^{d}$]  Energy discharged from time period $t$ to $t$+1 [MWh]
\item[\rule{0pt}{5pt}$q_{t}^{reg}$]  Energy allocated for frequency regulation from time $t$ to $t$+1 [MWh]
\item[\rule{0pt}{5pt}$s_{t}$] Energy storage state of charge at time period $t$+1 [MWh]

\end{IEEEdescription}}

\section{Introduction}
COVID-19 had significant effects on human life.
The most impactful year, that being 2020, proved to have a consequential effect on businesses.
Lockdown was one of the solutions to reduce the rapid spread of COVID-19.
Consequently, it reduced the energy demand used for running businesses, public utilities, entertainment activities, etc.
In addition, many people left the urban areas to move back home where they were able to work remotely and avoid the high living costs in larger cities.
As a result, the demand for electricity and the prices of fuel  were reduced, consequently suppressing  the wholesale electricity market prices \cite{ruan2020cross}. 

The revenue of energy storage systems (ESSs) comes from arbitrage and frequency regulation services \cite{NYISOstorage, kazemi2017operation, nguyen2019market}. 
Energy arbitrage oversees the purchase of energy when it is at a low price and/or low demand and sells it at a higher price and/or high demand.
Frequency regulation services are activities to maintain the electrical grid at a stable frequency of 60Hz \cite{walawalkar2007economics}.
ESSs are compensated for their services on stabilizing the grid frequency, i.e., 
they charge energy from the grid when the frequency is high and release their stored energy when the frequency is lower than the balance state.
The overall operations of an ESS can be modeled as a co-optimization problem in which the ESS aims to maximize its total revenue in both electricity and regulation markets subject to its physical constraints.  
However, they both depend on wholesale electricity prices, which were significantly suppressed in large population-dense cities by the reduction of electricity demand  induced by the business shut down and people migration. 

This paper focuses on the effects of COVID-19 on ESS operations \cite{ruan2020cross}, particularly in population-dense regions using a techno-economic approach. 
To this end,  
we consider the technology-specific properties of different energy storage technologies, namely Lithium-ion \cite{liu2015butler}, Advanced Lead Acid \cite{enos2014lead,  azzollini2018lead}, Vanadium Redox Flow \cite{lourenssen2019vanadium}, Lithium-Iron Phosphate \cite{swierczynski2013selection}, and Flywheel \cite{bolund2007flywheel}. 
We leverage the linear programming approach  to model the revenue maximization of the ESS in the electricity and regulation markets \cite{kazemi2017operation, nguyen2019market} using market data collected before, during, and after the COVID-19 pandemic \cite{concepcion2020quest}.
We conduct our analysis in the case study of New York City (NYC), the largest population with a high density of commercial infrastructure among cities in the United States, using  NYISO market data.  
Our analysis provides a comprehensive view of the impact of COVID-19 regard to the ESS revenues.

The rest of this paper is organized as follows.
Section II reviews key technology-specific properties of popular types of current energy storage technologies.
Section III discusses the electricity and regulation markets as well as the mathematical model of ESS' participation in these markets to maximize its revenue stream by arbitraging electricity trading and providing frequency regulation services.
Within this context, the techo-economic COVID-19 impact analysis for ESS is developed in Section IV. 
Section V presents the numerical results with a case study of NYC and Section VI concludes the paper. 



\section{Overview of Energy Storage Technologies}

\begin{figure}[http]
    \centering
	\includegraphics[width=0.95\textwidth]{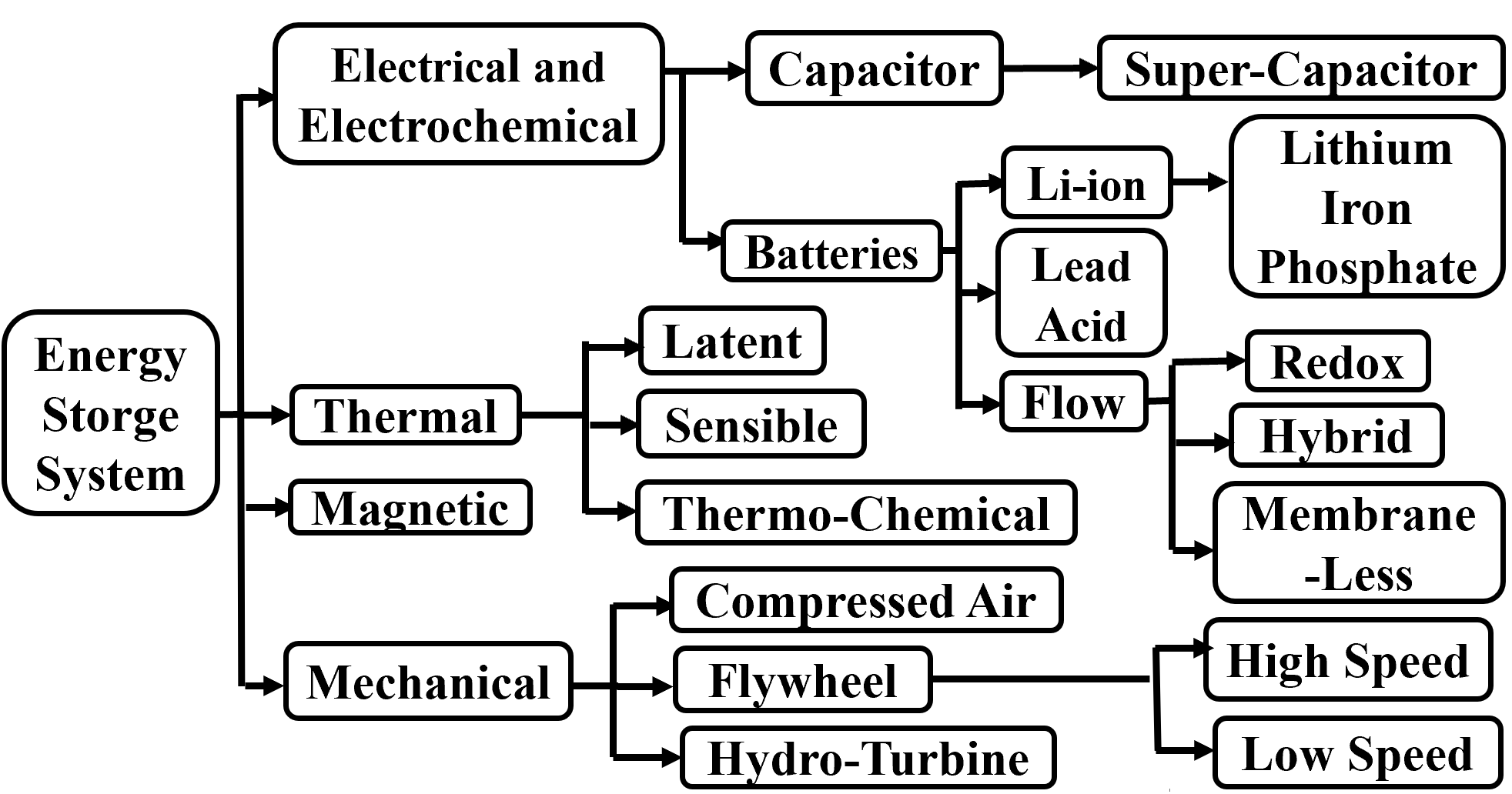}
	\vspace{-0.3cm}
	\caption{Taxonomy of Energy storage systems}
	\label{fig: ESS}       
\end{figure}


Figure \ref{fig: ESS} presents the overview of energy storage technologies with applications on the power grid.
Common energy storage technologies include Lithium-ion batteries, Advanced Lead Acid batteries, Vanadium Redox Flow Batteries, LFP batteries, and Flywheel.
The first four types are batteries, which are the most dominant energy storage form. 
Batteries can be dispatched quickly and provide power for long periods of time.
Consequently, batteries are effective to help offset the load and/or allow continuity during maintenance or outages of the grid.
Flywheels are not batteries.
However, they convert electrical energy into mechanical energy, thus storing energy similarly to traditional batteries at a lower cost.

\subsubsection{Lithium-ion battery}
Lithium-ion batteries are  constructed of an anode, cathode, separator, electrolyte, and two current collectors, one positive and the other negative where Lithium-ions (Li-ions) are the charge carriers.
The anode delivers Li-ions to the cathode, which initiates a flow of electrons between the two.
The majority of Li-ion batteries employ carbon materials in the anode, for example, graphite.
However, some utilize metal oxides like Lithium Titanate ($Li_{4}Ti_{5}O_{12}$), which is known for quick charging and having a high rate of chemical stability \cite{liu2015butler}. 
Li-ion batteries provide high energy and power density, making them capable of powering devices of all sizes or storing vast amounts of energy.
\subsubsection{Advanced Lead Acid battery}
Advanced Lead Acid batteries are assembled with two sheet lead plates, the anode, and cathode, that are in contact with two electrodes and a separator.
Lead acid batteries utilize the electrochemical conversion of the anode (lead metal) and the cathode (lead oxide) to Lead Sulfate.
The anode reacts with the Sulfuric acid to create Lead Sulfate, which acts as a reactant for the battery as well as a medium for ionic transport \cite{enos2014lead}. 
The separator is designed to prevent short circuits and protect the battery life span. 
The dominant separator used for advanced lead acid batteries is an absorbent glass mat (AGM) since AGM has low internal resistance and is capable of delivering high currents. 
It is well-established, cheap, safe, and is the most mature rechargeable storage technology \cite{azzollini2018lead}.
\subsubsection{Vanadium Redox Flow battery}
Vanadium Redox Flow Batteries (VRFBs) rely on electrochemical reactions to store and release energy whereas Vanadium refers to the element used in the battery and redox means reduction-oxidation.
VRFB stores the electrolytes in two separate tanks, away from the battery, and circulated in a loop with a pump.
The external surfaces of the loops contact each other at a membrane where the actual battery is.
The transfer of ions, particularly Vanadium, occurs at the membrane from the anode to the cathode \cite{lourenssen2019vanadium}.
The membrane’s purpose is to separate the positive and negative electrolytes while transferring the ions.
VRFB is often used for applications that require high power and energy density.

\subsubsection{Lithium-Iron Phosphate battery}
Lithium-Iron Phosphate battery (LFP) is a type of Li-ion battery.
The key difference, and where the name originates from, is the material of the cathode, i.e., $LiFePO_{4}$.  
LFP battery has a lower energy density, lower operating voltage (compared to other Li-ion batteries), low cost, high safety, and a long life cycle.
The increased safety is a byproduct of the batteries’ good thermal stability.
On its own, the LFP battery has low conductivity, which reduces its potential to have a high energy density.
This energy storage system is suitable for frequency regulation since it has a low cost of discharge/charge. 
Their implementation in affordable electric vehicles alongside it being a great choice to store electricity generated by photovoltaic cells further increases its value \cite{swierczynski2013selection}.

\subsubsection{Flywheel}
Flywheels are a form of energy storage that deviate from the rest of the previous ones discussed.
Unlike batteries, which store and release energy through electrochemical reactions, flywheels store and release energy kinetically.
A flywheel consists of a spinning rotor, a motor generator, a power electronics interface, bearings, and a housing.
The motor generator is generally integrated into the Flywheel and it is coupled directly to it.
Power electronics control the voltage and frequency created by the Flywheel.
This is done by rectifying the current and then converting it back to AC.
Filters are applied to reduce harmonics caused by high-frequency switching  \cite{bolund2007flywheel}.
The Flywheel store energy by rotating the mass at a high speed, adding more energy requires more speed\cite{subkhan2011new}. 
To discharge energy, it spins around an axis at a slower speed, which drives a motor at high speeds to convert mechanical energy to electrical energy.
Flywheel technology has been primarily used for transportation.
However, in recent times, it has been utilized for large-scale storage at the grid level because of its quick response, high energy density, stability, and low environmental impact characteristics \cite{fang2021methods}.

\begin{figure}[http]
    \centering
	\includegraphics[width=1\linewidth]{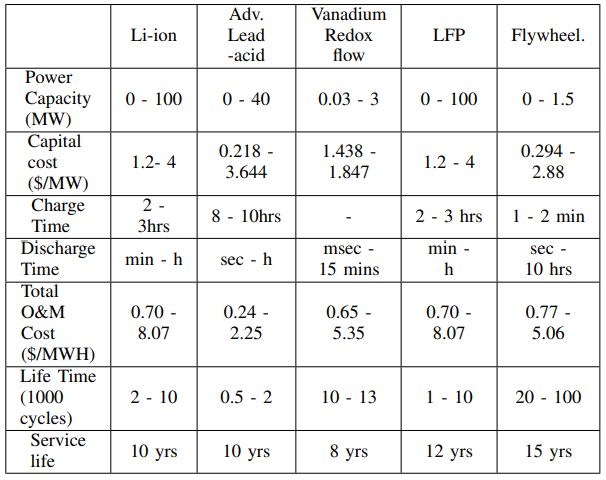}
	\vspace{-0.2 in}
	\captionof{table}{Characteristics comparison of ESS}
	\label{tab:ESS}    
\end{figure}
The key parameters of the five energy storage technologies are shown in Table \ref{tab:ESS}. 
Their values are based on \cite{concepcion2020quest}. 
Their values will affect the operations of ESS in the electricity and regulation markets, and consequently the obtained revenues.  

\section{Electricity and Regulation Markets with ESS' Participation}

In the United States, large-scale transmission grids are operated by Regional Transmission Organizations (RTOs) and Independent System Operators (ISOs). 
RTOs tend to cross state lines in terms of coverage.
ISOs, on the other hand, tend to cover only one state in terms of map coverage.
by the Federal Energy Regulatory Commission (FERC).
Both RTOs and ISOs are overseen by FERC \cite{lin2006economic}.
The goal of RTOs/ISOs is to ensure over time the 
efficient and reliable operation of the power transmission grid by running electricity  and regulation markets respectively. 
The operations of electricity markets are based on the two-settlement
system design \cite{nguyen2019market} in which power dispatches are determined in accordance with Locational Marginal Prices (LMPs), denoted as $\lambda_t$\footnote{An LMP determined for a particular time at a particular grid bus is the dual variable of the power balance constraint for this time and bus.}.
Additionally, the frequency regulation market is the market for ancillary services settled at the regulation price, e.g., Regulation Movement Price (RMP) in the case of NYISO,  
to maintain the constant frequency of the grid by second-to-second adjustment of power. 
Benefited from regulatory changes
such as FERC Order 755 which encouraged wholesale power
markets to compensate regulation resources based on the
actual regulation provided, and FERC Order 841 to remove barriers to the participation of ESS in the electricity and frequency regulation markets, ESSs can participate in both markets to increase their revenue streams \cite{NYISOstorage}.

\begin{figure}[http]
    \centering
	\includegraphics[width=0.95\textwidth]{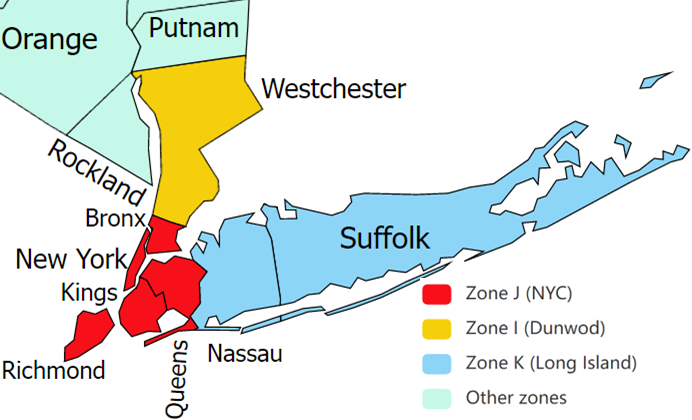}
	\vspace{-0.4pt}
	\caption{NYC areas and the corresponding zones in NYISO}
	\label{fig: NYC}       
\end{figure}

The ISO that manages the transmission grid in the state of New York is NYISO.
It consists of 11 transmission zones where NYC is zone J as shown in Figure \ref{fig: NYC} \cite{walawalkar2007economics}.
The NYISO was the first grid operator to develop market rules in accordance with FERC Order 755, allowing ESS to participate in wholesale markets as a regulation service provider. 
The NYISO 's effort to expand the role of storage through a full-market participation model was outlined in the 2017 State of Storage Report. That model allows grid operators and ESS to take better advantage of the capabilities that an ESS can provide to energy, capacity, and ancillary services markets.

\subsection{ESS' participation in the electricity market}

For ESSs to join the arbitrage business in NYISO, they must fulfill the minimum requirement of a discharge capability of four hours for market participation only in the electricity market \cite{walawalkar2007economics}.
The revenue in arbitrage of an ESS is the difference between the amount of money from selling/discharging the energy and buying/charging energy.
For simplicity, we assume that the charging, discharging, and other maintenance activity costs of the ESS are equal to zero.
Therefore, the ESS' revenue of arbitrage  from the period of time $t \rightarrow t+1 \in \mathcal{T}$ is  $\lambda_{t}(q_{t}^{d}-q_{t}^{r})$ where $\lambda_{t}$ denotes the electric price \$/MWh at time $t$, $q_{t}^r$ [MWh] denotes the charged energy and $q_{t}^d$ [MWh] denotes the discharged energy of an ESS during time $t$ to $t+1$.
The ESS' participation of in the electricity market by arbitraging energy to maximize its revenue can be modeled by the following optimization problem:
\begin{equation}
    \max\limits_{ q^{d},q^{r},s_t}\sum\limits_{t\in\mathcal{T}}\lambda_{t}(q_{t}^{d}-q_{t}^{r})
    \label{obj}
\end{equation}
subject to:
\begin{gather}
    s_{t+1}=\eta_{s}s_{t}+\eta_{c}q_{t}^{r}-q_{t}^{d},~~\forall t\in\mathcal{T}
    \label{con1}
\\
     0 \leq s_{t}\leq \overline{S},~~\forall t\in\mathcal{T}
    \label{con2}
\\
     q_{t}^{r}+q_{t}^{d}\leq \overline{Q},~~\forall i\in\mathcal{T}.
    \label{con3}
\end{gather}
The objective function (\ref{obj}) is to maximize the total revenue of ESS over the time horizon $\mathcal{T}$. 
Optimization variables include charged energy, discharged energy, and the state of charge of ESS at each time slot, i.e., $q^r_t, q^d_t, s_t$, subject to the following constraints. 
Constraint (\ref{con1}) captures the change of the ESS' state of charge, , $s_{t}$ [MWh], in two consecutive time slots, $t$ and $t+1$, based on the charged/discharged operations of ESS, $q^r_t$ and $q^d_t$.    
The parameter $\eta_{s}$[\%] represents the self-discharging efficiency of the ESS, which models the loss of the ESS' state of charge due to leakage.
Similarly, there is a difference between the amount of energy charged and the amount of energy that is actually stored after charging, which is captured by the round-trip efficiency, i.e., parameter $\eta_{c}[\%]$.
Furthermore, at anytime the $s_{t}$ of an ESS needs to be larger than or equal to zero and less than or equal to its energy capacity $\overline{S}$, which is captured in (\ref{con2}).
The amount of energy charged and discharged is also limited by the power rate $\overline{Q}$ of the ESS, which is captured in (\ref{con3}).

\subsection{ESS' participation in electricity and regulation markets}

\begin{figure}[http]
    \centering
	\includegraphics[width=1.0\columnwidth]{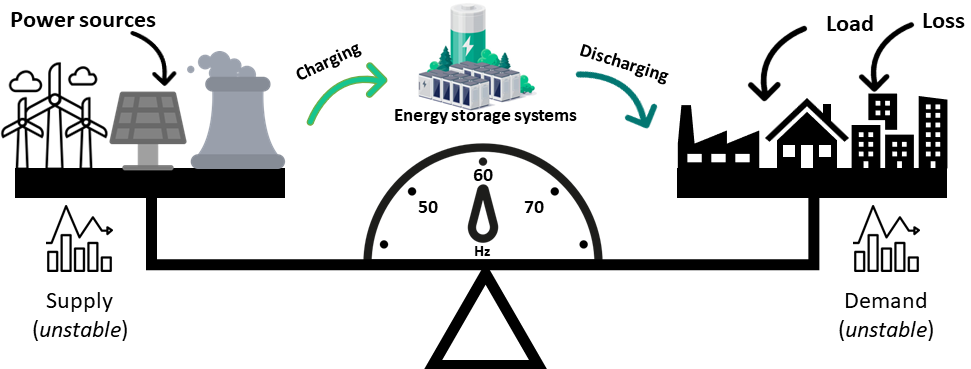}
	\vspace{-0.1 in}
	\caption{ESSs in frequency regulation services}
	\label{fig: frequency}       
\end{figure}

Figure \ref{fig: frequency} illustrates the use of ESS to provide frequency regulation services for continuously balancing generation and load as well as keeping the inter-area power exchanges at the scheduled values.
Due to the increasing penetration of intermittent renewable energy such as wind/solar and uncertain factors like weather and temperature, both the generation and demand sides of the power grid become uncertain, which can result in deviation of the grid frequency from its nominal value (i.e., 60Hz)
ESSs compensate for changes in frequency in two different ways.
When the frequency rises, i.e., lower demand and surplus of generation, 
ESSs offset the generation by storing energy (i.e charging up).
Conversely, when the frequency falls, i.e., higher demand and insufficient generation,
ESSs offset the imbalance by discharging energy.

In NYSIO, the automatic generation control (AGC) computes and allocates
the area control error (ACE)  to the scheduled regulation service suppliers every 6 seconds \cite{NYISOstorage}.
The allocation is proportional to the regulation movement (i.e., droop control) considering the physical limits and power grid constraints.
The performance of resources, e.g., ESS, in providing frequency regulation services is measured by the Performance Tracking System (PTS).
PTS index is estimated for each resource to adjust payments and charges based on how well the resource responds to the AGC command\cite{nguyen2019market}.  
The joint participation of ESS in the electricity and regulation markets can be modeled by the following optimization problem:
\begin{equation}
   \begin{split}
       \max\limits_{q^{d},q^{r},q^{reg}, s_t}\sum\limits_{t\in \mathcal{T}} & (\lambda_{t}(q_{t}^{d}-q_{t}^{r}+\delta_{t}^{ru}q_{t}^{reg}-\delta_{t}^{rd}q_{t}^{reg}) \label{obj2} \\
       & {+\lambda_{t}^{c}(q_{t}^{reg}-1.1(1-\gamma_{t})q_{t}^{reg}))e^{-Rt}}
   \end{split}
\end{equation}
subject to:
\begin{gather}
    {s_{t+1}}{=}{\eta_{s}s_{t}}{+}{\eta_{c}q_{t}^{r}}{-}{q_{t}^{d}}{+}{\eta_{c}}{\delta_{t}^{rd}}{q_{t}^{reg}}{-}{\delta_{t}^{ru}}{q_{t}^{reg}},~~\forall t \in \mathcal{T}, \label{soc2} \\
    0 \leq s_{t}\leq \overline{S},~~\forall t \in \mathcal{T}, \label{soc_lim2}\\
    q_{t}^{r}+q_{t}^{d}+q_{t}^{reg}\leq \overline{Q}, ~~\forall t \in \mathcal{T} \label{Q_lim2}\\
    r_{arb}  =\sum_{t\in\mathcal{T}}\lambda_{t}(q_{t}^{d}-q_{t}^{r}+\delta_{t}^{ru}q_{t}^{reg}-\delta_{t}^{rd}q_{t}^{reg}), 
    \nonumber \\
    r_{reg} =\sum_{t\in\mathcal{T}}\lambda_{t}^c(q_{t}^{reg}-1.1(1-\gamma_{t})q_{t}^{reg}). \label{con9}
\end{gather}


The objective function (\ref{obj2}) represents the total revenue obtained from both markets.  
In addition to charged and discharged energy, i.e., $q^r_t, q^c_t$, we need a new optimization variable, namely $q^{reg}_t$, to represent quantity bid in the frequency market. 
 During the operations, based on the AGC signal, a fraction of the capacity bid will be utilized for regulation down  and regulation up, which are captured by two parameters $\delta^{rd}_t$ and $\delta^{ru}_t$ respectively.
In particular, when the frequency is less than 60 Hz, the system sends a frequency-up signal and ESS discharges an additional amount of $\delta^{ru}_t q^{reg}_t$.
When the frequency is greater than 60 Hz, the system sends a frequency-down signal  and the ESS charges an additional amount of $\delta^{rd}_t q^{reg}_t$.
The instructed charged/discharged energy is added to the originally scheduled charged/discharged operations of ESS.
Consequently, the net energy exchange of ESS is $q_{t}^{d}-q_{t}^{r}+\delta_{t}^{ru}q_{t}^{reg}-\delta_{t}^{rd}q_{t}^{reg}$, which is arbitraged at the electricity price $\lambda_t$.  
The second line in (\ref{obj2}) represents the cash flow from the energy capacity which was offered in the Day-ahead market (DAM) contract subtracted from the penalty charge, 
Here,  $\lambda^c_i$ represents the regulation capacity price (RCP) and the constant 1.1 is the factor related to the penalty charge with NYISO (which could be different in other RTOs/ISOs), and $\gamma_{t}$ [\%] represents the NYISO tracked performance score on the ESS for a period of $t$ to $t+1$.
As a result, the actual energy storage supply from an ESS is the product of $\gamma_{t}$ and $q_{t}^{reg}$.
Finally, because the formulation is in the time domain, the inflationary factor $R$[\%] is included. 
Constraint (\ref{soc2}) represents the state of charge considering the scheduled charge/discharge and the instructed regulation up/down operations of the ESS.
Limits of the ESS state of charge are captured in (\ref{soc_lim2}).
The total energy exchange of the ESS is constrained by its charging capacity as in (\ref{Q_lim2}).
Finally, the revenue of arbitrage and the revenue of regulation services are calculated in (\ref{con9}) for records. 

\section{Techno-Economic COVID-19 Impact Analysis}
The noticeable impact of COVID-19 on NYSIO was a downward shift in electricity demand with a significant and almost-immediate drop in power consumption amongst residential, commercial, and industrial end-use sectors across the state of New York.
The situation was especially intense in NYC due to business shut-down and the migration of people.
In April 2020, when COVID-19 was rampant, the electricity demand in NYC was 14.1\% less compared to April 2019 \cite{ruan2020cross}.
Since electricity demand determines both the electricity supplied quantity and the marginal cost of dispatched energy, this sudden change in demand has disproportionately affected the fuel sources of marginal generating units, consequently alternating the electricity prices.

\begin{figure}[t!]
    \centering
	\includegraphics[width=1\textwidth]{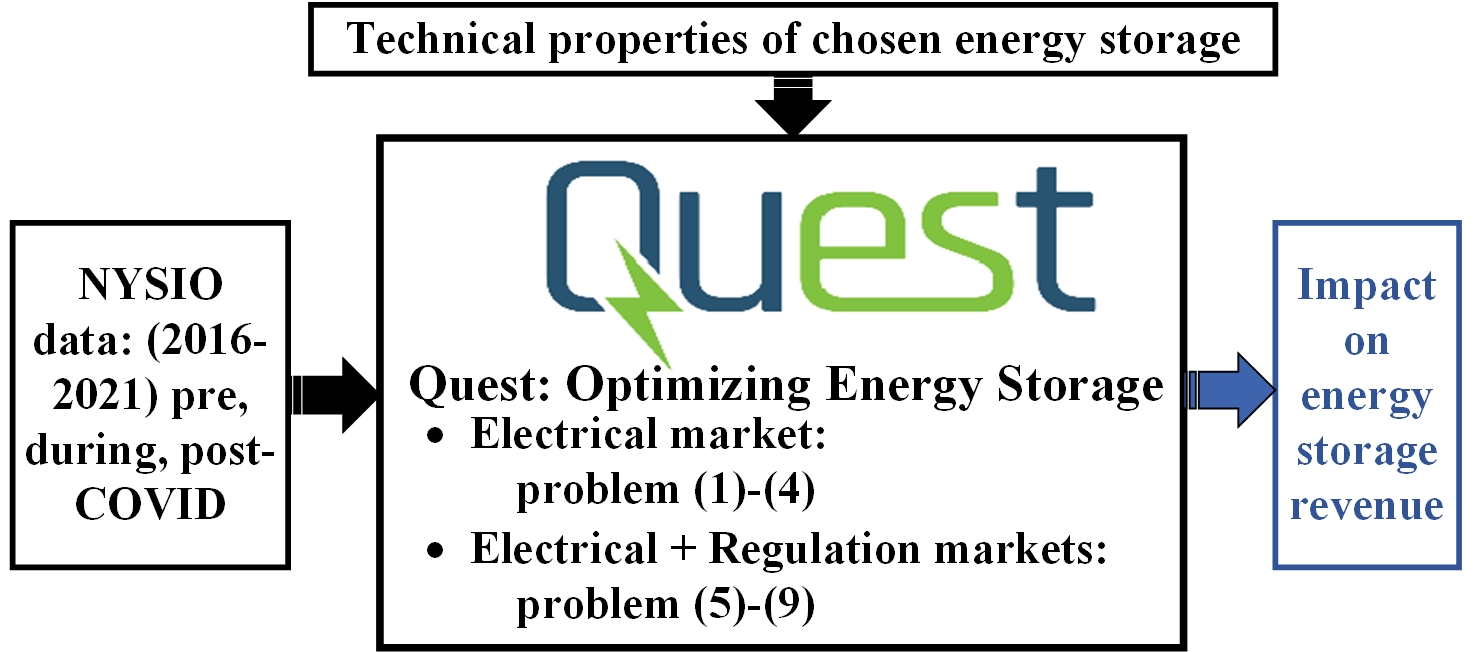}
	\vspace{-0.2 in}
	\caption{Techno-economic approach for COVID-19 impact analysis}
	\label{fig:Quest}       
\end{figure}

We propose a techno-economic approach for COVID-19 impact analysis as shown in Figure \ref{fig:Quest}. 
Mathematically, the problem (\ref{obj})-(\ref{con3}) and the problem  (\ref{obj2})-(\ref{con9})  are linear programs with two sets of parameters, the NYISO economic signals, (i.e., $\lambda_t$,  $\lambda^c_t$, $\delta^{ru}_t$/ $\delta^{rd}_t$) and ESS' physical parameters  (i.e., $\eta_s, \eta_c, \overline{S}, \overline{Q}$).
The first set was affected by COVID-19 through the changes in the NYISO electricity demand shift whereas the second set depends on the technical characteristic of the ESS technologies.
We run different operational modes of different ESS technologies in the NYC area using the NYISO data in pre, during, and post-COVID-19.
The time horizon of these problems is set for 24-hour daily operations.
We repeatedly solve problems (\ref{obj})-(\ref{con3}) and  (\ref{obj2})-(\ref{con9}) for all 365 days of the sequential years 2016-2021.  
We leverage the QuESt  package \cite{concepcion2020quest} developed by the Sandia National Lab to collect data and solve optimization problems with CPLEX as the linear programming solver. 
While NYC is selected for case studies, the framework can be applied for other U.S. regions.

\section{Numerical Results}

\begin{figure}[http]
    \centering
	\includegraphics[width=1\columnwidth]{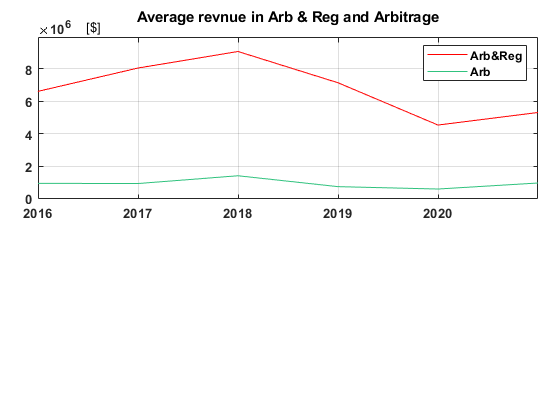}
	\vspace{-1.3in}
	\caption{Average revenue of ESSs installed in NYC from 2019-2021 under arbitrage and joint arbitrage \& regulation services modes}
	\label{fig4}       
\end{figure}

\begin{figure}[http]
    \centering
	\includegraphics[width=1\linewidth]{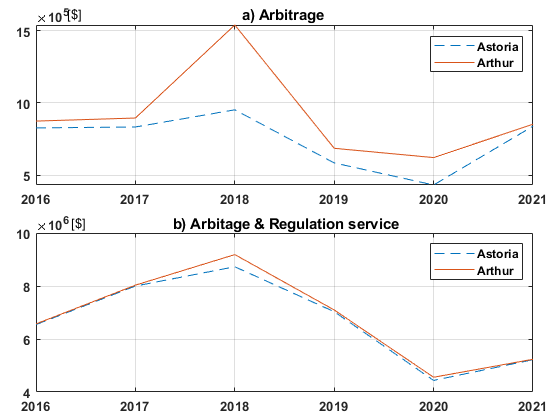}
	\vspace{-0.2 in}
	\caption{ Revenue of ESSs from 2016-2021 installed in 
Arthur and Astoria station  under a) Arbitrage and b) Joint arbitrage and regulation services (Arthur and Astoria are two electric substations in NYC)}
	\label{fig5}      
\end{figure}

\begin{figure}[t!]
    \centering
	\includegraphics[width=1\textwidth]{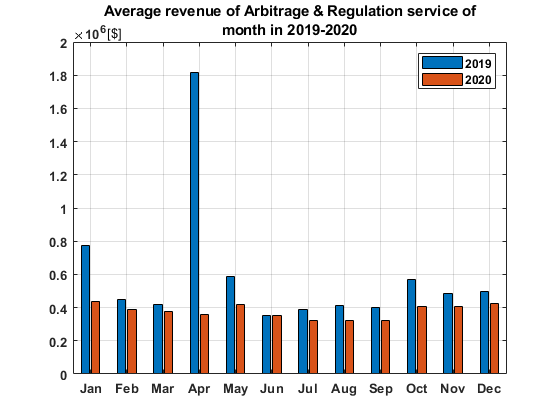}
	\vspace{-0.3 in}
	\caption{The average revenue of electric substations in arbitrage and regulation services during 12 months in 2019-2020.}
	\label{fig6}       
\end{figure}

\begin{figure}[http]
    \centering
	\includegraphics[width=1\textwidth]{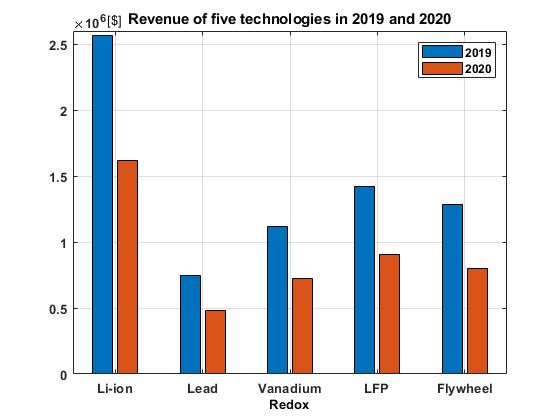}
	\vspace{-0.35 in}
	\caption{The change in average revenue of different types of energy storage  in pre and during COVID-19 (years 2019 and 2020).}
	\label{fig7}       
\end{figure}

We conduct our analysis with different ESS technologies (particularly Lithium-ion, Advanced Lead Acid, Vanadium Redox Flow, Lithium-Iron Phosphate, and Flywheel) whose parameters are summarized in Table \ref{tab2} \cite{concepcion2020quest}.
We consider cases where ESSs  are installed in Bronx, New York, Richmond, Kings, and Queens areas of NYC.
Their maximized revenues for operations in pre, during, and post-COVID-19 are collected and shown in Fig. \ref{fig4}-\ref{fig7}.

\begin{table}[http]
\centering
\caption{ESSs' parameters in formulation}
\begin{tabular}{|l|c|c|c|c|c|}
\hline
\multicolumn{1}{|c|}{} & Li-ion & \begin{tabular}[c]{@{}c@{}}Adv.\\ Lead\\ -acid\end{tabular} & \begin{tabular}[c]{@{}c@{}}Vanadium\\ Redox\\ flow\end{tabular} & \begin{tabular}[c]{@{}c@{}}LFP\end{tabular} & Flywhl. \\ \hline
\begin{tabular}[c]{@{}l@{}}\\\vspace{7pt}$\eta_{s}$ {[}\%{]}\end{tabular} & 100 & 100 & 100 & 100 & 100 \\ \hline
\begin{tabular}[c]{@{}l@{}}\\\vspace{7pt}$\eta_{c}$ {[}\%{]}\end{tabular} & 90 & 95 & 85 & 93 & 85 \\ \hline
\begin{tabular}[c]{@{}l@{}}\\\vspace{7pt}$\overline{S}${[}MWh{]}\end{tabular} & 24 & 7.5 & 60 & 7.8 & 5 \\ \hline
\begin{tabular}[c]{@{}l@{}}\\\vspace{7pt}$\overline{Q}$ {[}MW{]}\end{tabular} & 36 & 10 & 15 & 19.8 & 20 \\ \hline
\end{tabular}
\label{tab2}
\end{table}

Fig. \ref{fig4} demonstrates the significant impact of COVID-19, and shows that 2020 was the lowest year in average revenue of both operational modes, i.e.,  only arbitrage as well as arbitrage \& regulation services as well as.
However, the combination of arbitrage \& regulation services still yielded more revenue than the arbitrage itself.
In particular, by participating in both markets, the ESS can utilize better its flexibility for economic gains. 
The revenue increased in 2021 when the pandemic ends.
This is due to an increase in electricity demand, leading to more favorable economic incentives that the ESS can get from the NYISO markets.
Regardless, 2021 earnings remained low compared to pre-COVID-19 years since the overall economy was not yet fully recovered from the pandemic.

Fig. \ref{fig5} shows the impacts of the location on ESS revenue. 
Particularly, since the  NYISO employs an LMP mechanism for electricity prices, the ESS' revenue is also affected by the location in which it is installed.
There is a clear distinction between arbitrage \& regulation and arbitrage service. 
Arbitrage \& regulation were impacted greater than arbitrage as it has a much sharper negative slope in terms of revenue from the years 2019-2020.
This is because electricity demand has been unusually stable and low for all twelve months of 2020.

Fig. \ref{fig6} shows that the revenue of the arbitrage \& regulation services did not change much during the twelve months of 2020.
This is different from the magnitude of the change in revenue between the months for the normal year of 2019.
For the five ESS technologies, the impact of COVID-19 on their revenue was similar, albeit by different types or parameters.

Fig. \ref{fig7} shows that the revenue of the five technologies in arbitrage \& regulation services in 2020 were less than around 35-37\% in comparison to the revenue of 2019.
The result also reflects the difference in technical parameters between the five technologies in Table \ref{tab2}.
Li-ion battery has a high $\overline{S}$ and the highest $\overline{Q}$, which makes it more flexible in arbitrage \& regulation services.
As a result, before and during COVID-19, Lithium-ion technology had the highest revenue of the five technologies. 
Contrarily, Lead Acid has low $\overline{S}$ and lowest $\overline{Q}$, so its revenue is much lower than the others.

\section{Conclusion}
COVID-19’s adverse effects on the energy storage business in NYC were apparent and significant.
The usage of electricity fell along with the revenue of ESSs, producing the lowest year in terms of average revenue for both arbitrage \& regulation and arbitrage services.
Our techno-economic analysis with NYISO data reinforces these claims.
However, between the two, arbitrage services were affected more than arbitrage \& regulation services.
All the energy storage technologies were affected similarly in terms of revenue.
Our analysis gives a greater understanding of how pandemics cause harm to population-dense cities in terms of the power sector and how such disruption can financially affect the decarbonization effort, particularly the integration of energy storage. 
\bibliographystyle{IEEEtran}
\bibliography{ess}

\end{document}